\documentclass[aps,prc,twocolumn,]{revtex4}
\usepackage{graphics,graphicx,amsmath,amssymb,bm}

\newcommand{\be}{\begin{equation}}
\newcommand{\ee}{\end{equation}}
\newcommand{\bea}[1]{\begin{eqnarray}\label{#1}}
\newcommand{\eea}{\end{eqnarray}}

\newcommand{\lb}{\langle}
\newcommand{\rb}{\rangle}

\newcommand{\ket}[1]{|#1\rangle}
\newcommand{\matrEL}[3]{\langle#1|#2|#3\rangle}

\begin{document}

\title{Suppression of isoscalar pairing}

\author{G.F.~Bertsch and Simone Baroni}
\affiliation{Department of Physics and Institute for Nuclear Theory,
University of Washington, Seattle, WA 98195}


\begin{abstract}
The short-range nuclear attraction is stronger in the isoscalar
channel than in the isovector channel, as evidenced by the existence
of the deuteron and not the dineutron.  Nevertheless, apart from
light $N=Z$ nuclei, pairing is only seen in the isovector channel.
This is explained by the effect of the strong spin-orbit splitting
on the single-particle energies.  A semiquantitative argument is
presented treating the high-$j$ orbitals at the Fermi surface as
plane waves on a two-dimensional sheet.
\end{abstract}


\maketitle

Calculations with realistic interactions predict that in nuclear
matter pairing should be very strong in the $(S,T)=(1,0)$ channel
\cite{ba95}.  Nevertheless, pairing in nuclear structure is 
only visible in the $J=0$ isovector channel, except for a few
light odd-odd $N=Z$ nuclei, for which the ground state
has $J=1$. The reason is that the spin-orbit field affects the
pairing in the isoscalar channel more strongly than in the isovector
channel.  The effect is implicit in shell-model calculations such as
those of ref. \cite{po98}, dealing with the nonappearance  of $T=0$
pairing. Still, it is of interest to see how the effect comes about in 
a qualitative, analytic treatment of the interaction in $jj$ coupled
orbitals. We
first note that the high-$j$  orbitals are localized at the nuclear
surface, and their local momentum is mainly tangential to 
the surface.  In the limit of a very large nucleus, one can
treat the interactions as those of particles
localized on a 2-dimensional plane tangent to the surface.
On this plane, the Fermi surface is a circle and we need only
to investigate the interactions on the circle.  

\begin{figure}[!h]
  \includegraphics[width=0.28\textwidth]{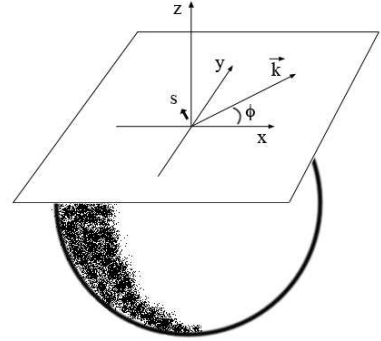}
  \caption{The geometry of high-j orbital state $\ket{\phi,\uparrow_p}$
  projected onto a tangent plane.  The directions of the spin $\vec s$
and linear momentum $\vec k$ correspond to a spherical orbital with
$j=\ell+1/2$. 
  \label{FIG:1}}
\end{figure}

We will analyze the momentum and spin of the particles in the vicinity
of the point $\vec r = (0,0,R)$ where $R$ is the radius of the nucleus.
The surface of the sphere is mapped onto a plane perpendicular to
the $z$ axis as shown in Fig. 1.  In this plane, the orbitals can 
be expressed in terms of their linear momentum $(k_x,k_y)$ and the
spin projection  $s_z$ as the kets  $ | k_x,k_y,s_z\rb$.   
In the presence of the spin-orbit field,
the eigenfunctions will be have the spin oriented in the $xy$ plane
and perpendicular to the momentum $\vec k$
in the $xy$-plane.  For the tangent plane shown in Fig. 1,
the preferred orientation corresponding to states with 
$j=\ell + 1/2$ is $\hat z  \times \hat k$.  Thus, the
preferred state with $k$ along the $x$-axis has spin oriented
along the $y$-axis, 
\be\label{EQ:1}
  |x \uparrow_y\rb={e^{i \pi / 4}\over \sqrt{2}} |k, 0, \uparrow_z\rb + 
    {e^{-i \pi / 4}\over \sqrt{2}} | k,0, \downarrow_z\rb.
\ee
In the general case,  when $\vec k$ is oriented at
an angle $\phi$ with respect to the $x$-axis, the preferred state is given by
\bea{EQ:2}
\lefteqn{
        \ket{\phi \uparrow_p}={\cal R}_\phi\ket{ x \uparrow_y}
        } \nonumber \\
  & = & {e^{i (\pi / 4+\phi/2)}\over \sqrt{2}} | k\cos \phi, k \sin \phi, \uparrow_z\rb +
        \nonumber \\
  &   & { e^{-i (\pi / 4+\phi/2)}\over \sqrt{2} } | k\cos \phi, k \sin \phi, \downarrow_z\rb
\eea
where the subscript on $\uparrow_p$ stands for the \emph{preferred}  axis
of spin orientation.  This is shown in Fig. \ref{FIG:1}.

\begin{figure}[!h]
\includegraphics[width=0.35\textwidth]{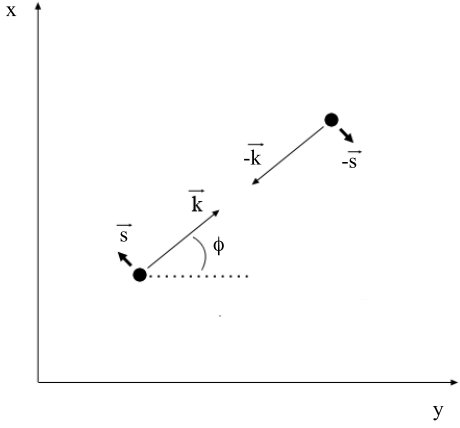}
  \caption{Visualization, of the state $\ket{2p,\phi,\uparrow_p \uparrow_p}$
  in a two-dimensional sheet, corresponding to two particles 
  on the nuclear surface, scattering in time reversal states, with 
  spin orientations dictated by the spin-orbit interaction.}
  \label{FIG:2}
\end{figure}

To treat the pairing interaction, we consider the set of 2-particle
states defined by the product of one-particle states having momenta $k$ and $-k$. 
The state with preferred spin orientation for both particles is
\be\label{EQ:3}
  \ket{2p,\phi,\uparrow_p \uparrow_p}=\ket{\phi\uparrow_p}\ket{(\phi+\pi)\uparrow_p}
\ee
as shown in Fig. \ref{FIG:2}.
We will later need to transform the spins to a common frame using
Eqs. (1) and (2).  In the usual $z$-oriented basis, the spin wave function of the
preferred 2-particle state is
\bea{EQ:11}
  \lefteqn{
           \ket{2p,\phi,\uparrow_p\uparrow_p}=
          } \nonumber \\ 
  & = & \frac{1}{2}\left(
                         e^{i\phi}\ket{\uparrow_z\uparrow_z}
                        +e^{-i\phi}\ket{\downarrow_z\downarrow_z}
                        \right. \nonumber \\
  & & \quad     \left. 
                        +e^{-i\pi/2}\ket{\uparrow_z\downarrow_z} 
                        +e^{i\pi/2}\ket{\downarrow_z\uparrow_z}
                   \right) \nonumber \\
  & = & \frac{1}{2}\left(
                         e^{i\phi}\ket{1,1}  
                        +e^{-i\phi}\ket{1,-1}
                        -i\sqrt{2}\ket{0,0}
                   \right). \nonumber \\
\eea
In the last line, we expressed the spin wave function  in the coupled form
$| S, S_z\rb$.

The interaction will mix wave functions having different orientations 
$\phi$, and the mixed state for spin $s_1,s_2$ will have the form
\be
 N \int_0^{2\pi} d \phi  f(\phi) |2p, \phi, s_1,s_2\rb
\ee
where $f(\phi)$ is an amplitude and $ N$ is a normalization factor.
Due to the rotational symmetry in
the $xy$ plane, we can take the amplitudes to have the form
$f(\phi)=e^{i M \phi }$.  Note that the wave function has a definite
particle exchange symmetry because the oriented function transforms
as  $P_{12} |2p,\phi, s_1 s_2\rb = - | 2p,\phi+\pi, s_1s_2\rb$.  Thus,
the even-$M$  mixed states are antisymmetric (ie. $T=1$) while the odd-$M$
states are symmetric ($T=0$).
We determine the norm $N$ by defining the overlap matrix element,
\be\label{EQ:4}
  \lb2p, \phi', s_1,s_2|2p, \phi, s_3,s_4\rb = \delta(\phi-\phi') \delta_{s_1,s_3}
  \delta_{s_2,s_4}.
\ee
Then the normalized wave functions for the interacting states are
\be\label{EQ:5}
  \ket{2p,M,s_1,s_2}={1\over \sqrt{2\pi}} \int^{2\pi}_0 d \phi\,e^{i M}  |2p, \phi, s_1,s_2
\rb.
\ee

We are now ready to calculate matrix elements.  
For a warm-up exercise, we start with a
spin-independent interaction  
\be
v={v_W\over 2 \pi}\delta^{(2)}(\vec r_1-\vec r_2)
\ee
and go back to the original spin representation, 
quantizing along the $z$-axis.  The matrix element is
\be\label{EQ:6}
  \lb 2p, \phi', s_{1z},s_{2z} | v | 2p, \phi, s_{3z},s_{4z}\rb = 
  {v_W\over 2 \pi}\delta_{s_{1z},s_{3z}}\delta_{s_{2z},s_{4z}}
\ee
where we have set the area corresponding to the integration over
the center-of-mass to one.  The matrix
element is independent
of $\phi$ and $\phi'$, which leads to a uniform mixing of $\phi$ values
in the paired state, ie. $M=0$ in the wave function Eq. (7).
The expectation value of the interaction energy in the $M=0$ state is 
given by
\be\label{EQ:7}
  \lb 2p, M=0,s_{1z},s_{2z}| v | 2p, M=0,s_{1z},s_{2z}\rb =v_W
\ee 
It is easy to include a spin dependence in the interaction in the
$s_z$ representation.
We  write the spin-dependent interaction as
\be\label{EQ:8}
v = {1\over 2 \pi}\left(v_0 P(0) + v_1 P(1)\right)
\ee
where $P(S)$ projects on the total spin $S$.  If our space includes all 
spin projections, i.e. the states $|2p, M=0,s_1,s_2\rb$ with the four 
orientations $s_1,s_2=\pm1$, we can recouple the spin states  
into a singlet and a triplet with 3 substates.  In that representation,
the interaction energy including the spin projector is $ v_0$ for
the singlet state and $ v_1$ for each of the three triplet states.  

Now we come to the main question, how are these interaction 
energies reduced when the space is restricted to the preferred 
spin orientation?
For this we need the matrix element of the interaction in the 
$\phi-$representation, Eq. (4),
\bea{EQ:12}
  \lefteqn{
  \matrEL{2p,\phi',\uparrow_p\uparrow_p}{v}{2p,\phi,\uparrow_p\uparrow_p} = } \nonumber \\
  & = &
  \frac{1}{4}\left(
                    e^{i(\phi-\phi')}\matrEL{1,1}{v}{1,1} 
             \right. \nonumber \\
  &   &      \quad \left.
                   +e^{-i(\phi-\phi')}\matrEL{1,-1}{v}{1,-1}
                   +2\matrEL{0,0}{v}{0,0}
             \right) \nonumber \\
  & = & 
  \frac{1}{8\pi}\left(
                    e^{i(\phi-\phi')}v_1
                   +e^{-i(\phi-\phi')}v_1
                   +2v_0
             \right). \nonumber \\
\eea
To find the eigenstates, we
apply the interaction to states of the form Eq. (7).
It is obvious that the three terms in Eq. (11) select
different $M$ values, with the $v_1$ interaction surviving for 
$M=\pm1$ and the $v_0$ interaction surviving for $M=0$.
The expectation values of the interaction energy are
\be
    \matrEL{2p,M,\uparrow_p\uparrow_p}{v}{2p,M,\uparrow_p\uparrow_p}=
\ee
$$  
   \frac{v_0 }{2} \,\,\,\, {\rm for}\,\,M=0,
$$
$$
   \frac{v_1}{4} \,\,\,\, {\rm for}\,\,M=\pm 1.
$$
Thus, the reduction from the interaction strength in the full 
spin space is a factor of two for the $S=0$ but a factor 4 for the
$S=1$.  As long as the intrinsic strength of the spin-triplet interaction is
not greater than twice the strength of the spin-singlet interaction, the
ordinary $J=0$ pairing will be favored within a single $j$-shell (for
$j$ large).  The factor of 2 suppression of the singlet pairing
can seen directly 
from the values of the appropriate $LS-jj$ recoupling coefficient.
For the shell with $j=\ell+1/2$, the relevant recoupling coefficient is
\be
\lb (jj)^{J=0} | (\ell \ell)^{L=0}, S=0\rb = \sqrt{2j+1\over 4 j}\rightarrow
\sqrt{1\over 2}.
\ee
The situation is somewhat more complicated for the triplet pairing
because both $L=0$ and $L=2$ couplings contribute to the single $j$-shell
interaction.  When both couplings are taken into account in a 
contact interaction, one finds that the suppression is indeed a factor
four in the limit of large $j$ \cite{Schwenk}.

\end{document}